\documentclass[12pt]{iopart}

\usepackage{graphicx}
\usepackage{color}
\usepackage{slashed}
\usepackage{epstopdf}

\begin{document}

\title[S-Z Su, Y-Y Zhao and X-J Wen]{Finite volume effects of the Nambu-Jona-Lasinio model with the running coupling constant}

\author{Shou-Zheng Su$^{1}$, Ye-Yin Zhao$^{2}$ and Xin-Jian Wen$^{3}$}

\address{$^{1}$Basic Experimental Teaching Center, Anhui Sanlian University, Hefei 230601, China}
\address{$^{2}$Sichuan University of Science and Engineering(SUSE), Zigong 643000, China}
\ead{yeyin.zhao@gmail.com}
\address{$^{3}$Institute of Theoretical Physics, Shanxi University, Taiyuan 030006, China}
\ead{zhengshsu@163.com}
\vspace{10pt}
\begin{indented}
\item[\today]
\end{indented}

\begin{abstract}
With the Schwinger's proper-time formalism of the Nambu-Jona-Lasinio model, we investigate the finite volume effects in the presence of magnetic fields. Since the coupling constant $G$ can be influenced by strong magnetic fields, the model is solved with a running coupling constant $G(B)$ which is fitted by
the lattice average $(\Sigma_u+\Sigma_d)/2$ and difference $\Sigma_u-\Sigma_d$. The investigation mainly focuses on the constituent quark mass and the thermal susceptibility depending on the magnetic fields, the temperatures and the finite sizes. For the model in finite or infinite volume, the magnetic fields can increase the constituent quark mass while the temperatures can decrease it inversely. There is a narrow range of the box length that makes the effects of finite volume perform prominently. The model will behave close to infinite volume limit for larger box length. It is shown that the influence of finite volume can be changed by magnetic fields and temperatures. Finally, we discuss the thermal susceptibility depending on the temperature in finite volume in the presence of magnetic fields.
\end{abstract}

%
%
%
%
%

\section{\label{sec:level1}Introduction}
The investigations of finite volume effects have great importance for the strongly interacting matter and attract many authors to enthusiastically do theoretical and experimental work \cite{a01}. The strongly interacting matter is essentially described by the Quantum ChromoDynamics(QCD) which correctly give most features of interactions between quarks and gluons. Experimentally the study of strongly interacting matter is mainly performed by the heavy-ion collisions at present. The strongly interacting matter produced in a heavy-ion collision always has finite volume which depends on the size of the colliding nuclei,the collision center of mass energy and the centrality of the collision \cite{a02}. The hadronic fireballs in relativistic nuclear collision reaction have volumes corresponding to a size of $2fm$ in radius or more \cite{a03}. The quark-gluon plasma(QGP) produced in high-energy heavy-ion collisions, which is also thought to have permeated the first microseconds of the Universe and to have cooled sufficiently to transform to hadronic matter soon, has sizes estimated between $2fm$ and $10fm$ \cite{a04,a05}. The fireballs of QGP formed in ultra-relativistic heavy-ion collisions undergo phase transition in a special range of temperature, volume and chemical potential \cite{a06,a07}. Therefore the volume plays an important role in the properties of strongly interacting matter produced in heavy-ion collisions.\par
The effects of finite volume have been thought over for decades in QCD and, especially, their analyses are encouraged by the simulations of QCD on finite, discrete Euclidean space-time lattices \cite{a01,a08}.  The finite volume effects appear sensibly until the box sizes up to $L\simeq5fm$ in the lattice simulations of light quark mass and strange quark mass \cite{a09}. Theoretically the investigation of finite volume effects has been completed by many effective approaches such as the Dyson-Schwinger equations of QCD  \cite{a10,a11,a12}, the quark-meson model \cite{a13,a14}, the non-interacting bag model \cite{a15}, the Nambu-Jona-Lasinio(NJL) model \cite{a16,a17,a18,a19}, the linear sigma model \cite{a20,a21} and others \cite{a22,a23,a24}. The boundary conditions should be imposed on the effective models when the strongly interacting matter constrained in finite volume. There are many results worked on antiperiodic boundary condition(APBC) as well as periodic boundary condition(PBC) \cite{a25,a26,a27,a28}, sice no restrictions impose on the spatial direction for finite size systems \cite{a19}. In Ref. \cite{a19}, the authors also put forward the application of stationary wave condition(SWC) considering that quark's wave function equals zero on the boundary. The effects of finite volume may be induced by the spherical MIT boundary condition when the strongly interacting matter is considered to be constrained in a sphere  \cite{a29}. By means of the Multiple Reflection Expansion(MRE) formalism, the finite volume effects can be taken into account in the Polyakov loop Nambu-Jona-Lasinio model and the deconfinement phase transition could be influenced by a finite radius \cite{a30}. In this work we adopt the well-known antiperiodic boundary condition(APBC).\par
The magnetic field has been shown by plenty of investigations that it has great influence on the thermodynamics and the phase transition of strongly interacting matter \cite{a31,a32}. In the presence of magnetic fields, the effects of finite volume occur consequently and the magnetic catalysis effect remains in all considered ranges of finite sizes  \cite{a33}. Furthermore the magnetic field is also confirmed that it has influence on the coupling constant \cite{a34}. Many authors have made efforts in constructing a magnetic-field-dependent running coupling constant \cite{a35,a36,a37,a38}. As the running coupling constant depends on magnetic fields, it will have important influence on the phase transition as well as the stability of quark matter \cite{a39,a40}. It is reasonable that we investigate finite volume effects in this work with a running coupling constant in the presence of magnetic fields.\par
In this paper, we will investigate finite volume effects of strongly interacting matter with the framework of the two-flavor NJL model. In the following content, we first show a general Schwinger's proper-time formalism of the NJL model in the presence of magnetic fields in Sec. \ref{sec:level2}. The finite temperatures are taken into account by applying the Matsubara formalism. In Sec. \ref{sec:level3}, the model is generalized to finite volume with the antiperiodic boundary condition. The running coupling constant depending on magnetic field is determined through fitting dimensionless quantities to the lattice results. By solving the gap equation in finite volume, we show the numerical results in Sec. \ref{sec:level4} and the results are compared with the cases of infinite volume. Finally, a short summary is given in Sec. \ref{sec:level5}.\par

\section{\label{sec:level2}Schwinger's Proper Time Formalism of the NJL model}
In the presence of an constant magnetic field, the two flavor Nambu-Jona-Lasinio model, which is an effective low-energy model for QCD and is superior to investigate quark matter at finite density or temperature, can be described by the Lagrangian
\begin{eqnarray}
\label{formula01}
\mathcal{L}_{NJL}=\bar{\psi}(i\slashed{D}-\hat m_c)\psi+G[(\bar{\psi}\psi)^2- (\bar{\psi}\gamma_5\vec{\tau}\psi)^2].
\end{eqnarray}
The two flavor quark field $\psi=(\psi_u,\psi_d)^T$, the current quark mass matrix is $\hat m_c=diag(m_u,m_d)$ and $G$ is the conpling constant. For simplicity we adopt $m_u=m_d=m_c$ with the isospin-symmetric limit. The covariant derivative $D^{\mu}=\partial^{\mu}+i\hat QA^{\mu}$ with the electric charge matrix $\hat Q=diag(2e/3,-e/3)$ in flavor space and $A^{\mu}$ is the electromagnetic gauge field. To investigate finite size effects within a constant magnetic background field $B$, we can choose the Landau gauge $A^{\mu}=(0,-By,0,0)$ which means the magnetic field along $z$ direction. In the mean-field approximation, the interaction terms are assumed to be deviated small from their thermal average and then the Lagrangian can be simplified as
\begin{eqnarray}
\label{formula02}
\mathcal{L}_{MF}=\bar{\psi}(i\slashed{D}-M)\psi+G\langle\bar{\psi}\psi\rangle^2,
\end{eqnarray}
where the constituent quark mass $M$ is self-consistently determined by the gap equation
\begin{eqnarray}
\label{formula03}
M=m_c-2G\langle\bar{\psi}\psi\rangle.
\end{eqnarray}
The thermal average fields $\langle\bar{\psi}\psi\rangle$ in this formula is called quark condensate. It can be defined by the trace of the dressed quark propagator
\begin{eqnarray}
\label{formula04}
\langle\bar{\psi}\psi\rangle=-\int\frac{d^4p}{(2\pi)^4}\Tr[iS(p)].
\end{eqnarray}
\par
The original purpose of the Schwinger's proper-time method is to maintain invariance properties in field calculations \cite{b01}. Then it is widely used to calculate higher loop and investigate the hadron and chiral phase transition as well as finite volume effects \cite{b02}. The Schwinger's proper-time method starts with the Green's function for the particle field
\begin{eqnarray}
\label{formula05}
(i\slashed{D}-M)S(x,y)=\delta(x,y).
\end{eqnarray}
In the coordinate space $S(x,y)$ and $\delta(x,y)$ can be regarded as the matrix elements of operators $\hat S$ and $\hat 1$ respectively. Consequently the Green's function can be expressed as
\begin{eqnarray}
\label{formula06}
\hat S=\frac{1}{i\slashed{D}-M}=\frac{-\slashed{D}+M}{-(\slashed{D})^2+M^2}=(-\slashed{D}+M)i\int_{0}^{\infty}dse^{-is(H-i\epsilon)},
\end{eqnarray}
\begin{eqnarray}
\label{formula07}
S(x,y)=\langle x|\hat S|y\rangle=(-\slashed{D}+M)i\int_{0}^{\infty}ds\langle x|e^{-is(H-i\epsilon)}|y\rangle,
\end{eqnarray}
where we have defined $H=-(\slashed{D})^2+M^2$. The meaningful idea of Schwinger's proper-time method is to consider $H$ as an Hamiltonian that describes evolution of some systems in proper time $s$. The state is defined as $|x(s)\rangle=e^{iHs}|x\rangle$. Then the matrix element of $e^{-iHs}$ can be viewed as transformation function from a state $|y(s=0)\rangle$ to another state $|x(s)\rangle$, i.e.
\begin{eqnarray}
\label{formula08}
\langle x|e^{-iHs}|y\rangle=\langle x(s)|y(0)\rangle.
\end{eqnarray}
The operators also depend upon the proper time parameter and evolute as the equation of motion in the Heisenberg Picture:
\begin{eqnarray}
\label{formula09}
i\frac{dx_{\mu}}{ds}=[x_{\mu},H]=-2i\Pi_{\mu},
\end{eqnarray}
\begin{eqnarray}
\label{formula10}
i\frac{d\Pi_{\mu}}{ds}=[\Pi_{\mu},H]=-2iq_{f}F_{\mu\nu}\Pi^{\mu}.
\end{eqnarray}
The transformation function can be solved from the differential equations
\begin{eqnarray}
\label{formula11}
i\frac{\partial\langle x(s)|y(0)\rangle}{ds}=\langle x(s)|H|y(0)\rangle,
\end{eqnarray}
\begin{eqnarray}
\label{formula12}
[i\frac{\partial}{\partial x^{\mu}}-q_{f}A_{\mu}(x)]\langle x(s)|y(0)\rangle=\langle x(s)|\Pi_{\mu}(s)|y(0)\rangle,
\end{eqnarray}
\begin{eqnarray}
\label{formula13}
[i\frac{\partial}{\partial y^{\mu}}-q_{f}A_{\mu}(y)]\langle x(s)|y(0)\rangle=\langle x(s)|\Pi_{\mu}(0)|y(0)\rangle,,
\end{eqnarray}
with the boundary condition
\begin{eqnarray}
\label{formula14}
\langle x(s)|y(0)\rangle|_{s\rightarrow 0}=\delta(x-y).
\end{eqnarray}
Following the details in Ref. \cite{b01,b03}, the transition function is finally expressed as
\begin{eqnarray}
\label{formula15}
\langle x(s)|y(0)\rangle=\frac{-i}{(4\pi s)^2}e^{-\frac{i}{4}(x-y)q_fF\coth(eFs)(x-y)-\frac{1}{2}Tr\ln\frac{\sinh(q_fFs)}{q_fFs}-is(\frac{q_f}{2}\sigma F+m^2)},
\end{eqnarray}
where the integration involved the Wilson line is neglected since it has no effect on the gap equations. Taking the Fourier transformation of $S(x,y)$ and carrying out the integration with respect to the coordinate variables, the quark propagator in momentum space is calculated as
\begin{eqnarray}
\label{formula16}
S(p)=&\int_{0}^{\infty}dse^{-is\{M^2-[(p^0)^2-(p^3)^2]+\frac{(p^1)^2+(p^2)^2}{q_fBs\cot(q_Bs)}\}}\nonumber\\&\times [M-\gamma^\mu p_\mu-(\gamma^1p_2-\gamma^2p_1)\tan(q_fBs)][1-\tan(q_fBs)\gamma^1\gamma^2].
\end{eqnarray}
\par
With the quark propagator Eq. (\ref{formula16}), the Schwinger's proper-time formalism of the NJL model can be constructed eventually. By taking the trace in the Dirac space, the flavor space and the color space, the quark condensate can be calculated as
\begin{eqnarray}
\label{formula17}
\langle\bar{\psi}\psi\rangle=-4MN_c\sum_{f=u}^{d}\int\frac{d^4p}{(2\pi)^4}\int_{0}^{\infty}dse^{-is\{M^2- [(p^0)^2-(p^3)^2]+\frac{(p^1)^2+(p^2)^2}{q_fBs\cot(q_fBs)}\}}
\end{eqnarray}
The integration of the quark condensate with respect to momentum carries out without difficulty using Gaussian integral. After transferring this expression into the Euclidean space by taking $s\rightarrow-i\tau$, the quark condensate is finally calculated as
\begin{eqnarray}
\label{formula18}
\langle\bar{\psi}\psi\rangle=-\frac{MN_c}{4\pi^2}\sum_{f=u}^{d}\int_{0}^{\infty}\frac{d\tau}{\tau} \frac{|q_fB|}{\tanh(|q_fB|\tau)} e^{-\tau M^2}
\end{eqnarray}
In order to take account of finite temperature, the integral over the four-dimensional momentum in Eq. (\ref{formula17}) should be replaced by the Matsubara formalism \cite{b04}, namely,
\begin{eqnarray}
\label{formula19}
\int\frac{d^4p}{(2\pi)^4}f(p)\rightarrow iT\sum_{n=-\infty}^{+\infty}\int\frac{d^3p}{(2\pi)^3}f(i\omega_n,\vec{p}).
\end{eqnarray}
The zeroth component of the momentum is discretized by the fermion Matsubara frequencies $p_0=i\omega_n=i(2n+1)\pi T$. As a result, the quark condensate in finite temperature is expressed as
\begin{eqnarray}
\label{formula20}
\langle\bar{\psi}\psi\rangle=-\frac{MN_c}{4\pi^2}\sum_{f=u}^{d}|q_fB|\int_{0}^{\infty}\frac{d\tau}{\tau} \frac{e^{-\tau M^2}}{\tanh(|q_fB|\tau)}\{1+2\sum_{n=1}^{+\infty}(-1)^ne^{-\frac{n^2}{4T^2\tau}}\},
\end{eqnarray}
where we have used the properties of the Jacobi's theta function
\begin{eqnarray}
\label{formula21}
\vartheta_3(z|x)=\sum_{n=-\infty}^{+\infty}e^{i\pi xn^2}e^{2niz}=1+2\sum_{n=1}^{+\infty}e^{i\pi xn^2}\cos(2nz).
\end{eqnarray}
\begin{eqnarray}
\label{formula22}
(-ix)^{\frac{1}{2}}\vartheta_3(z|x)=e^{-\frac{iz^2}{\pi x}}\vartheta_3(-\frac{z}{x}|-\frac{1}{x}).
\end{eqnarray}
\par
Since the NJL model is nonrenormalizable for its quadratic fermionic interaction, it is necessary to employ a regularization scheme to handle divergent integrals skillfully in the model. In this proper-time formalism we choose an ultraviolet cutoff $\Lambda$ to replace the lower limit of integrations in above equations, namely,
\begin{eqnarray}
\label{formula23}
\int_{0}^{+\infty}f(\tau)d\tau\rightarrow\int_{1/\Lambda^2}^{+\infty}f(\tau)d\tau.
\end{eqnarray}
Therefore the cutoff $\Lambda$, together with the current quark mass $m_c$ and the coupling strength $G$ are three parameters of the model which shoud be determined by the pion decay constant $f_\pi$, the pion mass $m_\pi$, and the quark condensate in vacuum. In this work we adopt the parameters to be $m_c=4.516MeV$, $\Lambda=1164.1MeV$, $G\Lambda=3.608$ in Ref. \cite{cn01} where $f_\pi=92.4MeV$, $m_\pi=138MeV$ and $-\langle\bar{\psi}\psi\rangle_0 ^{1/3}=260MeV$.
\par

\section{\label{sec:level3}Finite volume effects in the presence of magnetic field}
The Schwinger's proper-time formalism of the NJL model in Sec. \ref{sec:level2} describes systems in infinite volume. It is meaningful to generalize the previous results to the systems of finite volume since quark matter is always produced in restricted space regions by high-energy collision experiments \cite{a05}. Assuming that the system under consideration is restricted in a box with equal side lengths $L$, the quantum fields will satisfy boundary conditions leading to the rules
\begin{eqnarray}
\label{formula24}
\int\frac{d^3p}{(2\pi)^3}f(p_1,p_2,p_3)\rightarrow\frac{1}{L^3}\sum_{n_1=-\infty}^{+\infty}\sum_{n_2=-\infty}^{+\infty} \sum_{n_3=-\infty}^{+\infty}f(\omega_{n_1},\omega_{n_2},\omega_{n_3}),
\end{eqnarray}
which the momenta are discretized as
\begin{eqnarray}
\label{formula25}
p_i\rightarrow \omega_{n_i}=\frac{2\pi}{L}(n_i+\alpha),\quad n_i=0, \pm 1, \pm 2, ... .
\end{eqnarray}
The boundary condition is usually called antiperiodic boundary condition if $\alpha=1$ and periodic one if $\alpha=0$. In this work we adopt the antiperiodic boundary condition. Consequently, by replacing the momentum integration in Eq. (\ref{formula17}) with Eq. (\ref{formula19}), Eq. (\ref{formula24}) and Eq. (\ref{formula25}), the quark condensate at finite temperature and volume can be expressed as
\begin{eqnarray}
\label{formula26}
\langle\bar{\psi}\psi\rangle=&-\frac{2MN_c}{L^3}\sum_{f=u}^{d}\int_{1/\Lambda^2}^{\infty}\frac{d\tau}{\sqrt{\pi\tau}}e^{-\tau M^2}\{1+2\sum_{n=1}^{+\infty}(-1)^{n}e^{-\frac{n^2}{4T^2\tau}}\}\nonumber\\&\times \sum_{n_1=-\infty}^{+\infty}e^{-\frac{\tanh(q_fB\tau)}{q_fB}\omega_{n_1}^{2}} \sum_{n_2=-\infty}^{+\infty}e^{-\frac{\tanh(q_fB\tau)}{q_fB}\omega_{n_2}^{2}} \sum_{n_1=-\infty}^{+\infty}e^{-\tau\omega_{n_3}^{2}},
\end{eqnarray}
where the proper time has been transferred into the Euclidean space and the ultraviolet cutoff is taken on.\par
\begin{figure}[!hbt]
\centering
\includegraphics[width=9.0cm]{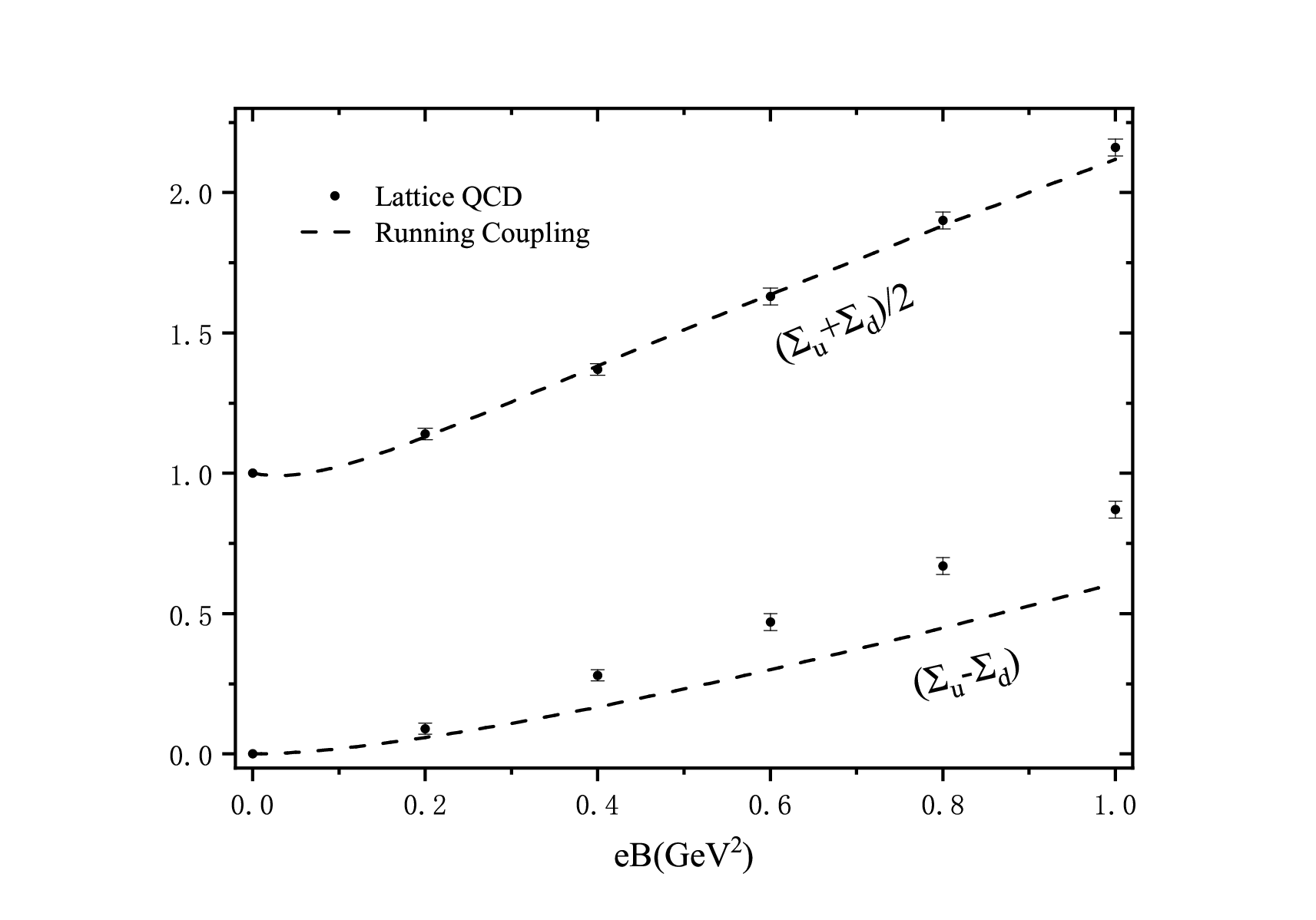}
\caption{The average $(\Sigma_u+\Sigma_d)/2$ and the difference $\Sigma_u-\Sigma_d$ at $T=0MeV$ are fitted to the lattice results of Ref. \cite{cn08}. The reasonable results are obtained with $\alpha=2.39$ and $\beta=0.002515$ in the running coupling constant Eq. (\ref{formula27}).}
\label{fig:01}
\end{figure}

As the quark condensate is given by Eq. (\ref{formula26}) in the presence of constant magnetic field, the gap equation Eq. (\ref{formula03}) can be solved with the fixed three parameters $\Lambda$, $m_c$ and $G$. The coupling constant $G$ controls the strength of strongly interaction in QCD and, however, can be influenced by sufficiently strong magnetic fields \cite{a33,cn02}. As a consequence, a appropriate form of running coupling constant depending on magnetic fields has been attempted by many authors to fit the lattice results \cite{a36,a37,a38,cn07}. A magnetic-field-dependent running coupling constant has influence on the constituent quark mass as well as the phase transition of the model \cite{a39,cn07}. In this work we adopt the magnetic-field-dependent running coupling constant \cite{a36}
\begin{eqnarray}
\label{formula27}
G(B)=\frac{G}{1+\alpha\ln(1+\beta\frac{eB}{\Lambda_{QCD}^2})},
\end{eqnarray}
where $\Lambda_{QCD}^{2}=200MeV$. The free parameters $\alpha$ and $\beta$ are fixed to get reasonable results of the lattice average $(\Sigma_u+\Sigma_d)/2$ for $T=0MeV$. The lattice average $(\Sigma_u+\Sigma_d)/2$ relates to quark condensate in NJL model by defining the dimensionless quantity \cite{cn08}
\begin{eqnarray}
\label{formula28}
\Sigma_{f}(B,T)=\frac{2m_c}{m_{\pi}^{2}f_{\pi}^{2}}[\langle\bar{\psi}_{f}\psi_{f}\rangle(B,T)- \langle\bar{\psi}_{f}\psi_{f}\rangle(0,0)]+1
\end{eqnarray}
In figure \ref{fig:01}, by solving the model in infinite volume with the running coupling constant Eq. (\ref{formula27}), the average $(\Sigma_u+\Sigma_d)/2$ as well as the difference $\Sigma_u-\Sigma_d$ at $T=0MeV$ is fitted to the lattice results of Ref. \cite{cn08}. The reasonable results are obtained with $\alpha=2.39$ and $\beta=0.002515$.
\par

\section{\label{sec:level4}Numerical results}
With the Schwinger's proper-time formalism, the model in the background of magnetic fields is extended to finite volume with antiperiodic boundary condition at finite temperatures. As the parameters $\alpha$ and $\beta$ in the running coupling constant  Eq. (\ref{formula27}) are fixed with the lattice data, the model can be presented by solving the gap equation Eq. (\ref{formula03}). In this section we will concentrate on the finite volume effects with the running coupling constant  Eq. (\ref{formula27}) in the presence of magnetic fields.\par

\begin{figure}[!hbt]
\centering
\includegraphics[width=9.0cm]{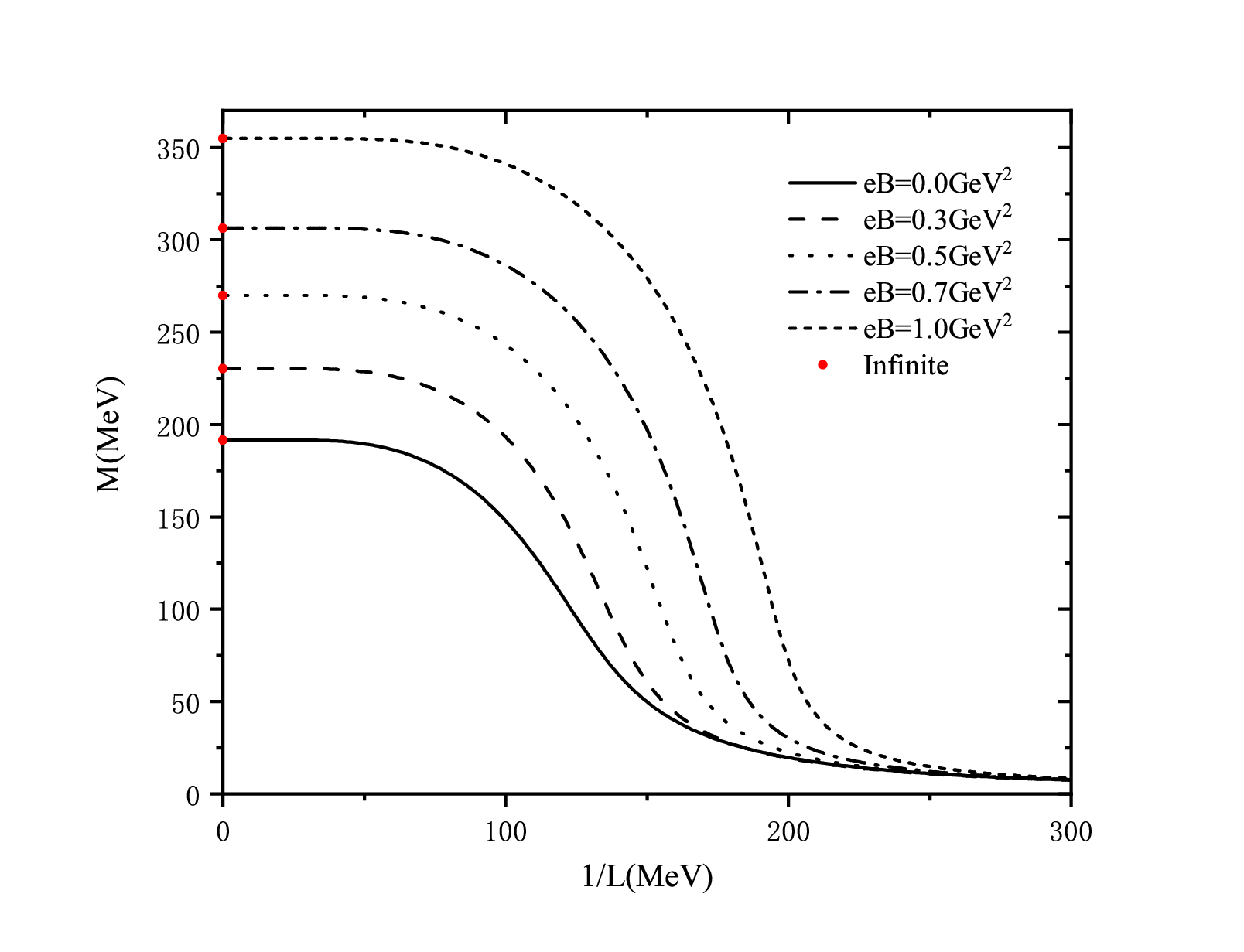}
\caption{The constituent quark mass $M$ at vanishing temperature depends on the inverse length $1/L$ in the presence of magnetic fields. The values at $1/L=0MeV$ correspond to ?those in infinite volume.}
\label{fig:02}
\end{figure}

When the quark condensate is calculated in finite volume, the constituent quark mass $M$ solved from the gap equation Eq. (\ref{formula03}) will be influenced apparently by finite volume. In figure \ref{fig:02}, the constituent quark mass at vanishing temperature is presented as a function of the inverse length $1/L$ in the presence of magnetic fields $eB=0.0GeV^2$, $eB=0.3GeV^2$, $eB=0.5GeV^2$, $eB=0.7GeV^2$, $eB=1.0GeV^2$. The results at $1/L=0MeV$ correspond to the cases in infinite volume. As the magnetic field becomes stronger, the constituent quark mass is obviously increased especially when the box length is close to infinite volume. While the increase falls off as the box length becomes smaller. When the box length is quite small, for example $1/L=250MeV$ which corresponds to $L\simeq0.79fm$, the constituent quark mass is small enough to close to chiral limit. As the box length $L$ increases away from quite small values, the constituent mass $M$ will sharply increases for all cases of $eB$ until it is close to the infinite volume limit. $M$ is close to the infinite volume limit at about $L=15fm$ for the case of $eB=0.0GeV^2$ and about $L=8fm$ for $eB=1.0GeV^2$. The box length that is close to the infinite volume limit is reduced by stronger magnetic fields.\par

\begin{figure}[!hbt]
\centering
\includegraphics[width=9.0cm]{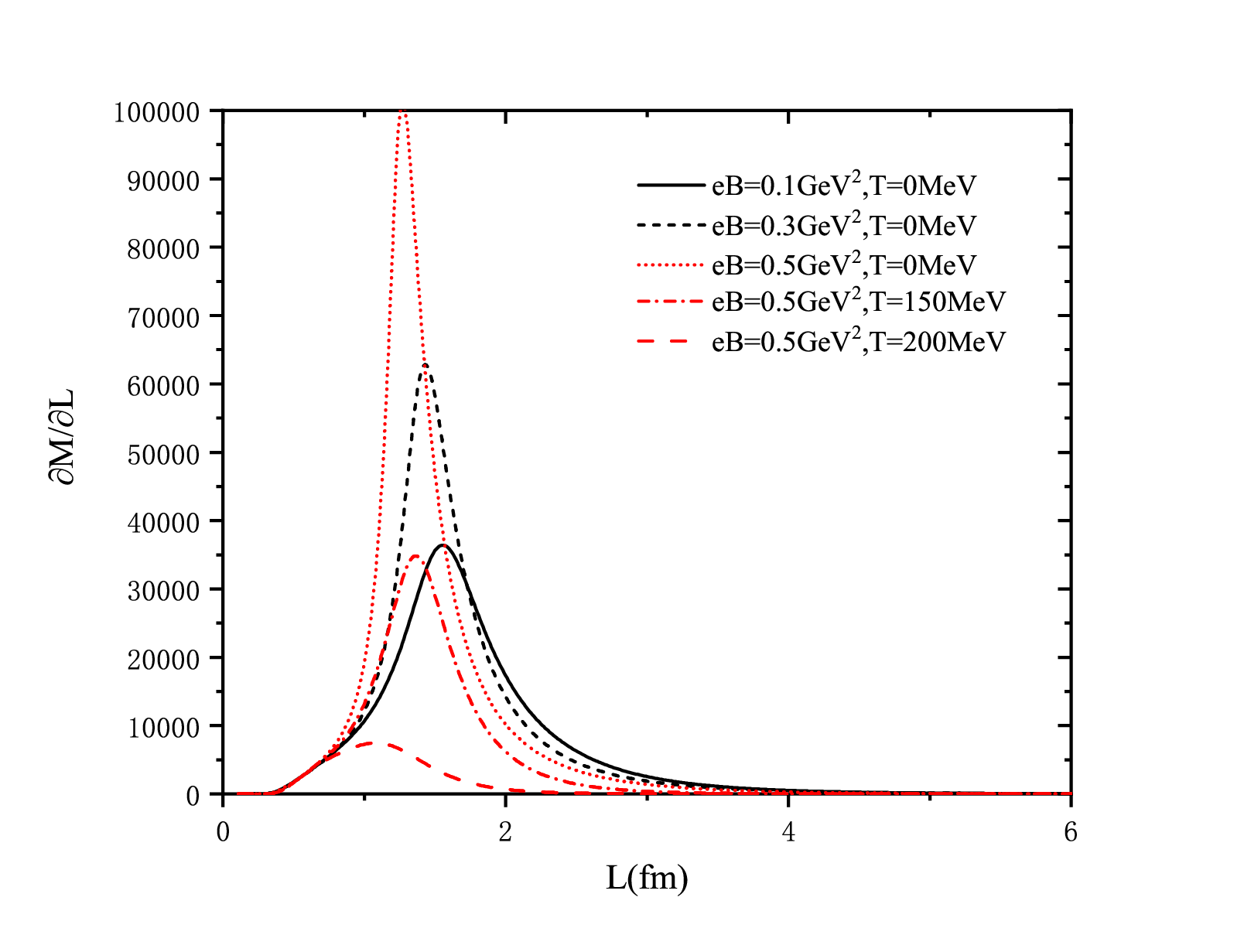}
\caption{The partial derivative respecting to length $\partial M/\partial L$ depends on the box length $L$ in the presence of magnetic fields $eB=0.1GeV^2$,$0.3GeV^2$ and $0.5GeV^2$. The temperatures are appropriately selected as $T=0MeV$,$150MeV$ and $200MeV$ for investigation.}
\label{fig:03}
\end{figure}

To find out the range of the box length which sharply affects $M$, the partial derivative respecting to length $\partial M/\partial L$ is presented in figure \ref{fig:03}. Overall, the constituent quark mass $M$ varies sharply with the box length when $\partial M/\partial L$ is massively greater than $0$. According to the cases of $T=0MeV$ at $eB=0.1GeV^2$, $0.3GeV^2$ and $0.5GeV^2$, the constituent quark mass $M$ will varies more sharply at stronger magnetic field. While it will varies less sharply at higher temperature according to the cases of $eB=0.5GeV^2$ at $T=0MeV$, $150MeV$ and $200MeV$. In addition, the box length that is close to the infinite volume limit can also be reduced by higher temperatures. The location of protrusions in figure \ref{fig:03} means that the constituent quark mass $M$ varies sharply for most cases when the box length is restricted in a narrow range between about $0.5fm$ and $4fm$. The narrow range can be reduced by stronger magnetic fields and by higher temperatures. It means that the finite volume effects could be weakened when the system is in stronger magnetic fields and higher temperatures.\par

\begin{figure}[!hbt]
\centering
\includegraphics[width=9.0cm]{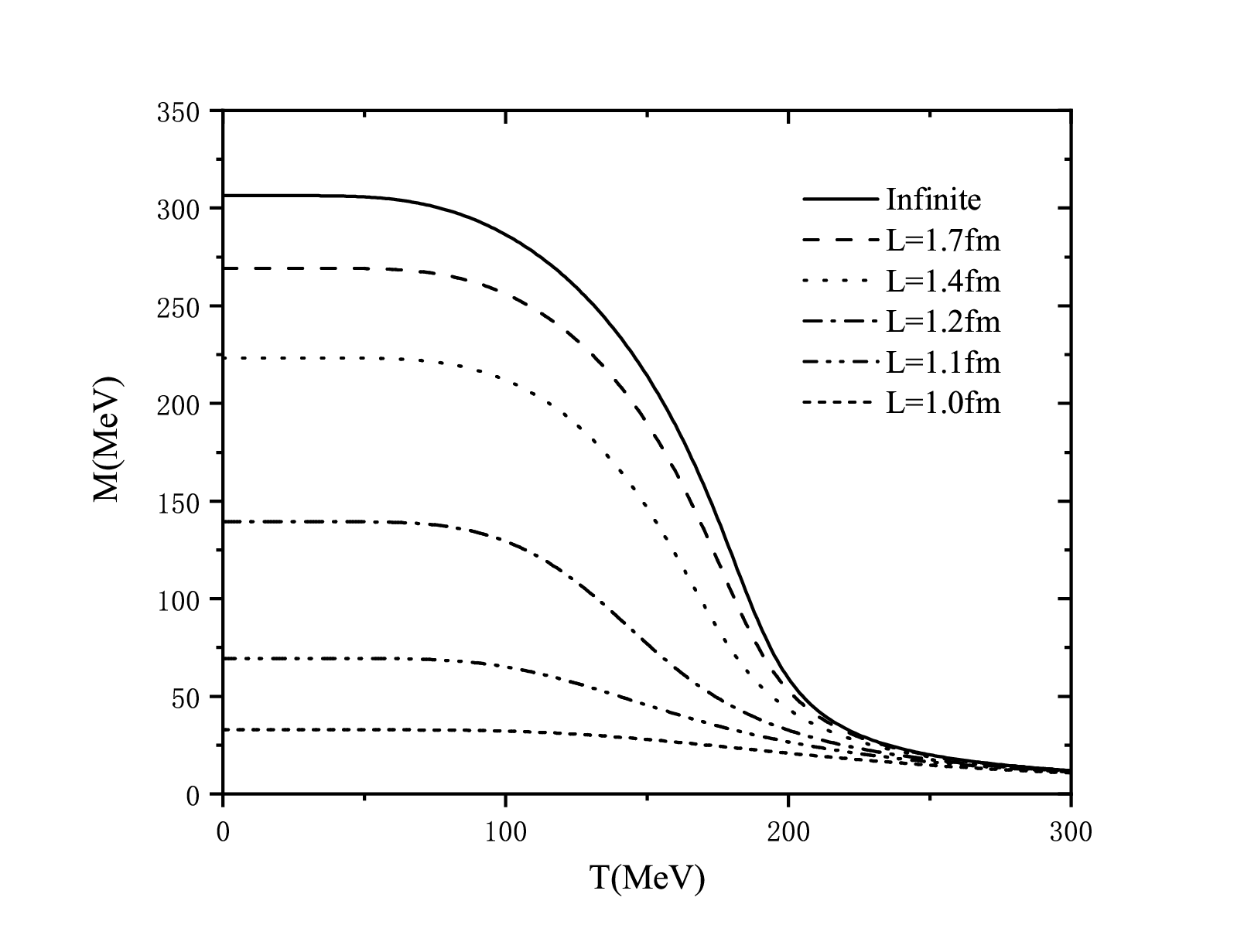}
\caption{The constituent quark mass $M$ depends on the temperature $T$ in the presence of magnetic field $eB=0.7GeV^2$ at the length $L=1.7fm$, $1.4fm$, $1.2fm$, $1.1fm$ and $1.0fm$. The solid line stands for the results in infinite volume.}
\label{fig:04}
\end{figure}

Therefore, the box length in the following numerical results is mainly set in the narrow range where the effects of finite volume appear obviously. In figure \ref{fig:04}, the constituent quark mass $M$ is presented as a function of magnetic field $eB$ at $L=1.7fm$, $1.4fm$, $1.2fm$, $1.1fm$ and $1.0fm$. For larger values of length, the behavior of $M$ is close to the line of infinite volume. Evidently $M$ decreases as the temperature increases as usual for all cases. While it decreases as the length decreases especially when the temperature is not high enough. When the temperature becomes high enough, for example $T=250MeV$, $M$ will decreases slightly by the decrease of $L$.\par

\begin{figure}[!hbt]
\centering
\includegraphics[width=9.0cm]{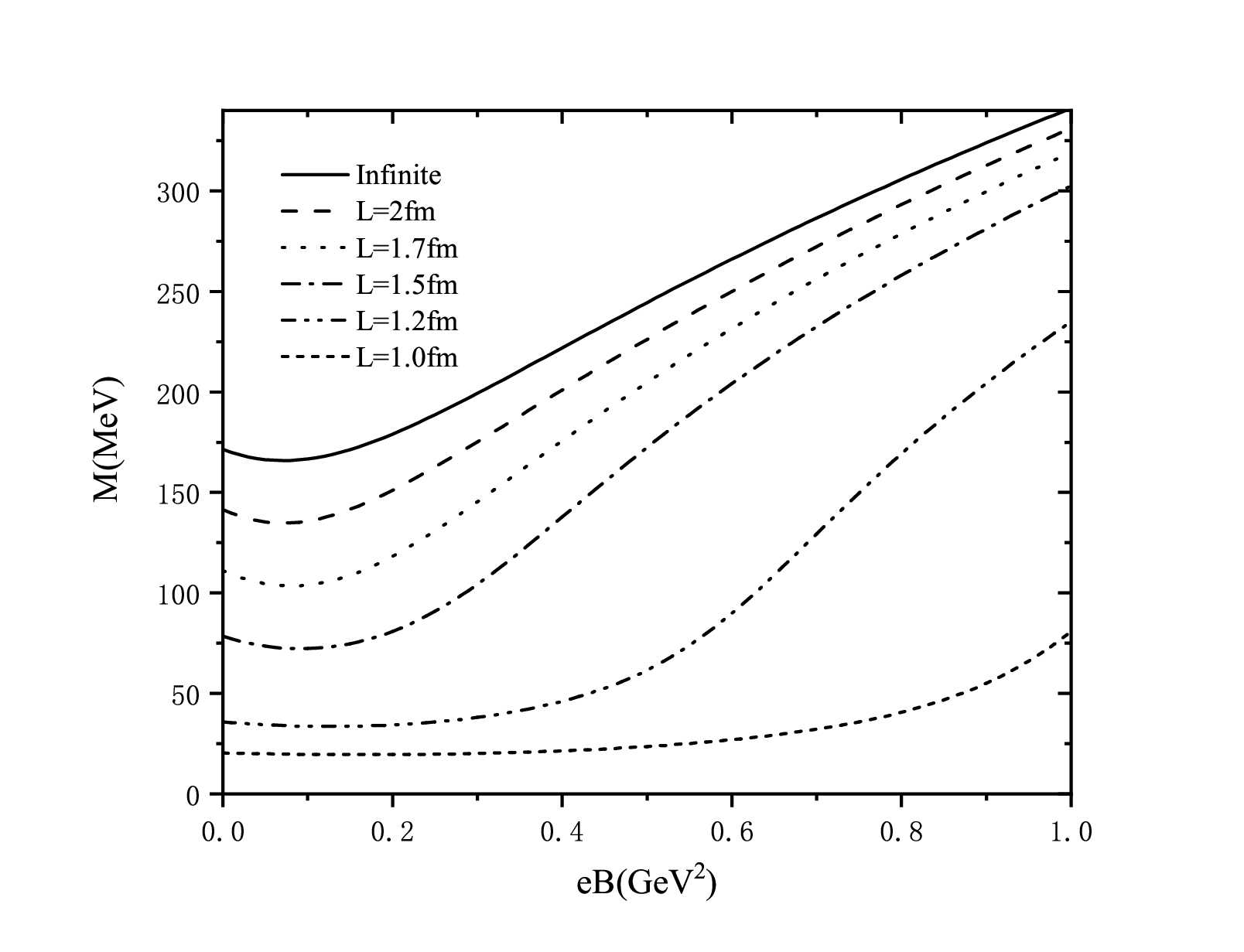}
\caption{The constituent quark mass $M$ depends on the magnetic field $eB$ at the box length $L=2fm$, $1.7fm$, $1.5fm$, $1.2fm$ and $1.0fm$ when the temperature $T=100MeV$. The solid line stands for the results in infinite volume.}
\label{fig:05}
\end{figure}

The dependence of constituent quark mass $M$ on the magnetic field $eB$ is presented in figure \ref{fig:05} at $L=2fm$, $1.7fm$, $1.5fm$, $1.2fm$ and $1.0fm$ when the temperature is $T=100MeV$. For larger values of length, the behavior of $M$ is also close to the line of infinite volume. For the system in infinite volume, $M$ will decreases as $eB$ increases beginning with $0GeV^2$. Soon afterwards $M$ will apparently increases by the increase of the magnetic field. Contrasting the lines of finite volume with the infinite volume, the constituent quark mass $M$ can be significantly decreased by the box length.\par

\begin{figure}[!hbt]
\centering
\includegraphics[width=9.0cm]{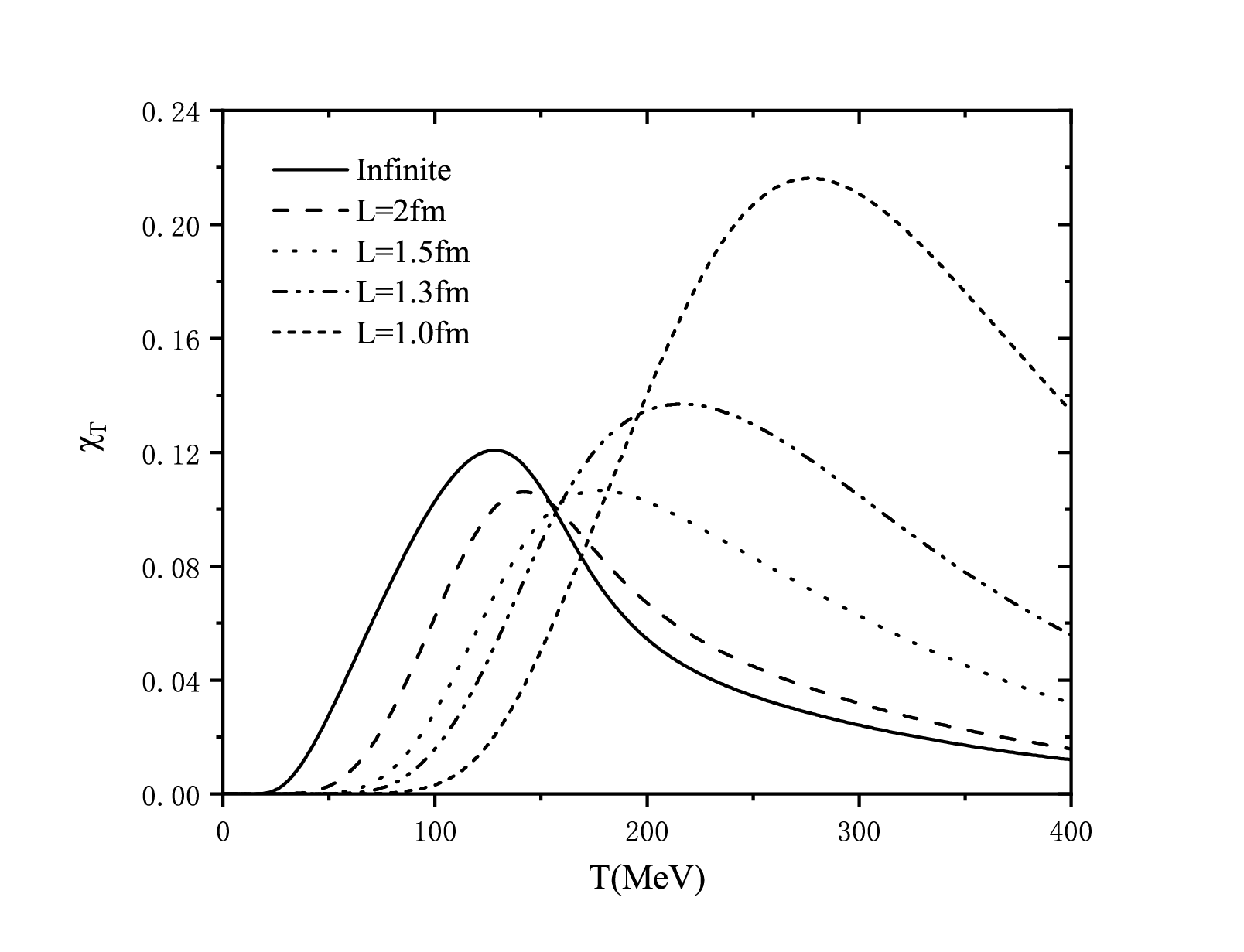}
\caption{The thermal susceptibility $\chi_{T}$ depends on the temperature $T$ in the presence of magnetic field $eB=0.1GeV^2$ at the length $L=2fm$, $1.5fm$, $1.3fm$ and $1.0fm$. The solid line stands for the results in infinite volume.}
\label{fig:06}
\end{figure}

The thermal susceptibility is defined as
\begin{eqnarray}
\label{formula29}
\chi_{T}=-m_{\pi}\frac{\partial\sigma}{\partial T},
\end{eqnarray}
where $\sigma$ is given by
\begin{eqnarray}
\label{formula30}
\sigma=\frac{\langle\bar{\psi}_{u}\psi_{u}\rangle(B,T)+ \langle\bar{\psi}_{d}\psi_{d}\rangle(B,T)}{\langle\bar{\psi}_{u}\psi_{u}\rangle(B,0)+ \langle\bar{\psi}_{d}\psi_{d}\rangle(B,0)}.
\end{eqnarray}
In figure \ref{fig:06}, the thermal susceptibility $\chi_{T}$ is presented as a function of temperature $T$ in the presence of magnetic field $eB=0.1GeV^2$ at the box length $L=2fm$, $1.5fm$, $1.3fm$ and $1.0fm$. Similarly the lines of $\chi_{T}$ is also close to the infinite volume for larger values of the length. As the box length decreases to small values showed in figure \ref{fig:06}, $\chi_{T}$ will move to the right, which leads to the peaks of the lines lying at higher temperatures. The peak of the thermal susceptibility $\chi_{T}$ defines a pseudocritical temperature. \par

\section{\label{sec:level5}Summary}
In this work we investigate the effects of finite volume at finite temperatures in the presence of magnetic fields with the Schwinger's proper-time formalism of the NJL model. The system in finite volume is considered by the antiperiodic boundary condition. Since the coupling constant $G$ can be influenced by sufficiently strong magnetic fields, the investigations of this paper work with the running coupling constant Eq. (\ref{formula27}) depending on the magnetic field. The running coupling constant Eq. (\ref{formula27}) is properly determined by fitting the free parameters with the lattice average $(\Sigma_u+\Sigma_d)/2$ and difference $\Sigma_u-\Sigma_d$ at $T=0MeV$. The numerical results of the model in finite volume are presented by solving the gap equation.\par
The magnetic field has the effect that increases the constituent quark mass of the model in both infinite and finite volume according to the dependence of $M$ on the inverse length. However, the increase falls off as the box length becomes smaller. When the length is quite small, the constituent quark mass is small enough to close to chiral limit. When the value of box length is appropriately larger, the constituent quark mass depending on the length will behave close to the infinite volume limit. Especially the box length that is close to the infinite volume limit can be reduced by stronger magnetic fields and by higher temperatures. For the system in finite temperatures and volume, there is a narrow range of the box length that makes the effects of finite volume perform prominently. This narrow range can also be reduced by stronger magnetic fields and by higher temperatures.\par
At finite temperatures, the constituent quark mass decreases by higher temperatures for both infinite and finite volume in the presence of magnetic fields. When the temperature is not very high, the constituent quark mass can decreases obviously as the box length decreases in the narrow range which effects of finite volume perform prominently. While the constituent quark mass will decrease slightly when the temperature is high enough. Considering the dependence of $M$ on the magnetic field at finite temperature, the constituent quark mass can also decrease obviously as the length decreases in the narrow range which effects of finite volume perform prominently. For the constituent quark mass depending on the temperature and the magnetic field, $M$ will behave close to the infinite volume limit when the box length is appropriate large. While $M$ has small values when the box length is quite small.\par
The thermal susceptibility $\chi_{T}$ changes obviously by the effects of finite volume. In the presence of magnetic fields, $\chi_{T}$ also behave close to the infinite volume limit when the length has appropriately large values. The diagram of $\chi_{T}$ will move to the right in the $T$-axis when the box length decreases in the narrow range which effects of finite volume perform prominently. Consequently the peaks of the $\chi_{T}$, which define pseudocritical temperature, move to higher temperatures.\par

\ack The authors would like to thank support from the National Natural Science Foundation of China (under the Grant Nos. 11875181, 11705163, 12275102 and 11475110), the National Key Research and Development Program of China (under the Grant No. 2022YFA1604900) and the Natural Science Foundation of Anhui Sanlian University (under the Grant Nos. KJZD2021003, KJZD2022010, KJZD2023007).

\end{document}